\documentclass[12pt]{article}

\newcommand{\be}{\begin{equation}}
\newcommand{\ee}{\end{equation}}
\newcommand{\bea}{\begin{eqnarray}}
\newcommand{\eea}{\end{eqnarray}}
\newcommand{\ti}{\tilde}

\font\mybb=msbm10 at 12pt

\def\bb#1{\hbox{\mybb#1}}

\def\P{\bb{P}}

\def\N{{\cal N}}
\def\id{\protect{{1 \kern-.28em {\rm l}}}}

\def\F{{\cal F}}

\def\dash{${-\!\!\!-\,\,}$}

\def\qnc{{Q_{N_c}^\alpha}}
\def\tqnc{{\tilde Q}_{N_c}^\alpha}
\def\qnc1{Q_{N_c+1}^\alpha}
\def\tqnc1{{\tilde Q}_{N_c+1}^\alpha}

\def\appendix#1{
  \addtocounter{section}{1}
  \setcounter{equation}{0}
  \renewcommand{\thesection}{\Alph{section}}
  \section*{Appendix \thesection\protect\indent \parbox[t]{11.715cm} {#1} }
  \addcontentsline{toc}{section}{Appendix \thesection\ \ \ #1}
  }

\makeatletter
\renewcommand\section{\@startsection {section}{1}{\z@}%
                                                                      {-3.5ex \@plus -1ex \@minus -.2ex}%
                                   {2.3ex \@plus.2ex}%
                                   {\normalfont\large\bfseries}}
\renewcommand\subsection{\@startsection{subsection}{2}{\z@}%
                                   {-3.25ex\@plus -1ex \@minus -.2ex}%
                                   {1.5ex \@plus .2ex}%
                                   {\normalfont\normalsize\bfseries}}
\makeatother

\newcommand{\Tr}{\mathop{{\rm Tr}}}

\begin{document}

\begin{titlepage}

\hfill\hbox to 3cm {\parbox{4cm}{
UCB-PTH-03/05 \\
LBNL-52343    \\
UCLA/03/TEP/5 \\
hep-th/0303115
}\hss}

\vspace*{5mm}
\begin{center}

\mbox{\hskip-.8em\large\bf Matrix Model Description of
Baryonic Deformations \hss}

\vspace*{8mm}

{Iosif Bena$^a$,~ Hitoshi Murayama$^b$,~ Radu Roiban$^c$,~ Radu Tatar$^d$}
\vspace*{5mm}


{\it ${}^a$ Department of Physics and Astronomy}\\
{\it University of California, Los Angeles, CA 90095}

\medskip
\smallskip

{\it ${}^{b,\,d}$ Department of Physics, 366 Le Conte Hall}\\
{\it University of California, Berkeley, CA 94720}\\

\vspace*{1mm}

{\it and}\\

\vspace*{1mm}

{\it Lawrence Berkeley National Laboratory}\\
{\it Berkeley, CA 94720}\\

\medskip
\smallskip

{\it ${}^c$Department of Physics}\\
{\it University of California, Santa Barbara, CA 93106}

\vspace*{3mm}


\end{center}


\begin{abstract}
We investigate supersymmetric QCD with $N_c+1$ flavors using an extension of the
recently proposed relation between gauge theories and matrix models. The
impressive agreement between the two sides provides a beautiful confirmation
of the extension of the gauge theory-matrix model relation to this case.
\end{abstract}

\vfill
\noindent March 2003
\vskip 1em
\hrule width10em
\vskip .3em
\noindent\hskip .5em
{\small E-mail:
\parbox[t]{8cm}{\tt ${}^a$ iosif@physics.ucla.edu \\
${}^b$ murayama@lbl.gov \\
${}^c$ radu@vulcan2.physics.ucsb.edu \\
${}^d$ rtatar@socrates.berkeley.edu }}

\end{titlepage}

\tableofcontents

\section{Introduction}

The fact that topological string amplitudes are closely related to certain holomorphic
quantities in the physical superstring theories was known for some time. A practical
incarnation of this relation is the recently discovered \cite{dv1,dv2,dv3} gauge
theory - matrix model connection. By now, this idea has been investigated
using three rather independent approaches.

In the original approach of \cite{dv1,dv2,dv3}, one starts from the open/closed string
duality implied by a geometric transition, and computes gauge theory superpotentials
using the fluxes of the dual geometry \cite{vafa}-\cite{cali}. The result is expressed in
terms of the partition function of a certain closed topological string theory. On the open
string side of the duality one relates the terms of the effective superpotential with
the partition function of the holomorphic Chern-Simons theory (which describes the
fields on the wrapped D-branes sourcing the geometry). Dijkgraaf and Vafa conjectured
that the open and closed partition functions are identical. Since the computation of the
open string partition function reduces to the computation of the partition function of a
large $N$ matrix model, this conjecture implies a certain relation between gauge
theory effective superpotentials and matrix models \cite{dv1}. This relation
was further strengthened by the study of the underlying geometry of the matrix model
and of the gauge theory \cite{dv2,dv3}.

In the second approach \cite{grisaru}, the effective glueball superpotential of an
$\N=1$ theory with adjoint matter was evaluated using superspace techniques. It
was found that only zero momentum planar diagrams contribute to this superpotential,
thus validating the original conjecture.

In the third approach
\cite{csw,wit1} the generalized Konishi anomalies of the field theory were used to
obtain relations between the generators of the chiral ring of the theory. These relations,
which under certain identifications can be reproduced from a matrix model,
can then be used to construct the effective superpotential.

Perhaps the most immediate extension of the matrix model-gauge theory
relation is to
theories with fields transforming in matter representations of the gauge group
\cite{Goteborg}-\cite{alishahiha}. For theories with fields transforming in the fundamental
representation of the gauge group it was suggested that the addition to the original DV
proposal involves matrix model diagrams with a single boundary. For arbitrary
generalized Yukawa couplings and a simple superpotential for the adjoint field this
proposal was proved in \cite{Bena:2002tn}. Furthermore, it was shown in
\cite{seib} that the matrix model fully captures the holomorphic
physics of theories with $N_f<N_c$, regardless of the complexity of the tree level
superpotential for the adjoint field.

Extending the correspondence to gauge theories with baryons turned out to be somewhat
more challenging. In particular, baryons only exist for theories with certain relations
between $N_f$ and $N_c$; therefore, taking the large $N_c$ limit to restrict to planar
diagrams is not possible. Moreover, the notion of boundary in diagrams with
baryons is not well-defined unless certain manipulations are performed.

In \cite{brr} an extension of the DV proposal to theories with baryons
was formulated and it was shown that for an  $SU(N_c)$ theory with $N_f = N_c$
quarks this proposal reproduces exactly the known gauge theory
physics. \footnote{Related work has appeared in \cite{Argurio}, where a  perturbative
field theory computation in the spirit of \cite{grisaru} was used to recover the
terms linear in baryon sources in the gauge theory effective superpotential.}
$$~$$

The goal of this paper is to extend the correspondence to supersymmetric QCD
with $N_f=N_c+1$ flavors. As it is well known \cite{seiberg}, this theory has $N_f$
baryons, $N_f^2$ mesons and a dynamically generated superpotential
\be
W={1 \over \Lambda^{2 N_c-1}}(B_i M^i_j \ti B^j - \det M)~~.
\ee
In order to relate this theory with the corresponding matrix model, we deform it with
the appropriate sources, and integrate out the mesons and baryons. We then
compare the result with the one given by matrix model planar one-(generalized)
boundary and find perfect agreement.

Our interest in theories with baryons has several reasons. The first is that neither of
the three routes by which the original matrix model-gauge theory relation was reached
seems easily extendable to these theories. \footnote{In particular, as we will see in the
last section of this paper, understanding the baryons at $N_f=N_c+1$ in geometric
transitions is quite difficult.} Second, in theories with only chiral flavors,
baryons are
the only objects one can construct. If one is to extend the matrix model-gauge theory
relation to such theories, understanding the r\^ole and the correct treatment
of baryons is crucial.

Last but not least, the superpotentials of theories similar to SQCD with
$N_f=N_c+1$ flavors\footnote{These are the so-called s-confining theories.
They have a description in terms of gauge-invariant composites everywhere
on the moduli space, and the effective superpotential for the confined degrees
of freedom is not singular at the origin of the moduli space.(For
discussions on s-confining theories see \cite{csa1,csa2} and
references therein).} are used as starting points for the construction of low
energy effective superpotentials in many theories
where symmetries do not determine these superpotentials directly. It is therefore
important to have a direct method of computing them.

One of the ways in which the validity of the $N_f=N_c+1$ superpotential is usually
tested is by obtaining the correct $N_f=N_c$ superpotential in the absence of baryonic
sources. However, this superpotential contains much more information, which can only
be captured by turning on {\it all} the baryon sources. The fact that the matrix model
reproduces the rather involved effective superpotential obtained with all the baryon
sources turned on is a very powerful confirmation of the validity of the
extension of the matrix model-gauge theory relation to baryons.

Integrating out all the fields in both the gauge theory and the matrix model is rather
complicated, and often results in rather unedifying expressions involving roots of
large degree polynomials. Fortunately, there exists a procedure \cite{JADE} which
allows us to compare gauge theory and matrix model results in theories containing
only mesons. Thus, to compare the gauge theory and the matrix model results
it is enough to  integrate out only two flavor fields on the gauge theory side
 (sections \ref{slick} and \ref{hard}), and to relate the result to the matrix 
model free energy obtained by treating the $N_c-1$ massless flavors as background
fields (section \ref{mm_gen_2}).

Except for $SU(2)$, where computing the values of the effective superpotential
at its critical points is not too difficult (sections 2 and 3), we will
only be comparing matrix and gauge theory results at $N_f=N_c-1$. In section
\ref{mm_all} we will derive these values, using gauge theory techniques,
and discuss a method for computing the full matrix model free energy.
As an example we  apply this method to the case of an $SU(2)$ gauge theory and recover
all expected results, already discussed in sections \ref{su2_ft} and \ref{mm2}.

Before proceeding, let us remark that in our case there is no distinction
between the unitary  matrix model and the hermitian one. This is due to the fact that
we will be interested in theories containing only fields transforming in the fundamental
representation. Since these fields are not constructed out of generators of the gauge
group, they are the same both in the $SU(N_c)$ and in the  $U(N_c)$ theories.
Thus, the matrix integral is the same for both gauge groups.

To fix the notation, Latin indices from the beginning of the alphabet ($a,b,c$) are
$SU(N_c)$ indices; Latin indices from the middle of the alphabet ($i,j,k$) are 
$SU(N_f=N_c+1)$ indices; Greek indices from the beginning of the alphabet ($\alpha,\beta,\gamma$)
are $SU(2)$ indices corresponding
to two flavor fields which are singled out. They take the values $N_c$ and $N_c+1$.
Hatted Latin indices from the middle of the alphabet are $SU(N_f=N_c-1)$ indices,
corresponding to the flavor symmetry unbroken by the quark masses (but nevertheless
broken by the presence of baryon operators).


\section{Review of the Dijkgraaf-Vafa proposal for flavors \label{generalities}}

In a series of papers \cite{dv1,dv2,dv3}, Dijkgraaf and Vafa
proposed a perturbative method for computing the effective glueball
superpotential of certain $\N=1$ theories with fields
transforming in the adjoint and bifundamental representations of the gauge group.
According to this proposal, the planar diagram contribution to the free energy of a
certain matrix model yields the effective superpotential of the corresponding $\N=1$
gauge theory. In 't~Hooft's double line notations these diagrams have the topology
of a sphere.

When fields transforming in the fundamental representation of the gauge
group (quarks) are present one must also include the free energy arising from
planar diagrams with one boundary (diagrams with the topology of a
disk) \cite{Goteborg,rr}. More explicitly, the gauge theory effective superpotential
is given by
\be
W_{\rm {eff}}(S,\Lambda)=N_cS(1-\ln{S\over \Lambda^3}) +
N_c{\partial \F_{\chi=2}\over \partial S}
+N_f \F_{\chi=1}~~,
\label{DV}
\ee
where the first two terms are also present in a theory with only adjoints, and
the third term is the contribution of the flavors.

If baryonic sources are also added, the diagrams that can be constructed become
more complicated. However, only the planar diagrams with as many index
loops as a diagram with one boundary contribute to the effective
superpotential \cite{brr}. Since the number of colors
and the number of flavors are related, it is not possible to select the
relevant diagrams by taking the limit in which the number of colors is large.
Thus, the planar diagrams with baryon sources have to be
selected by hand. In \cite{brr} it was shown that planar baryonic diagrams
for $SU(N_c)$ theories with $N_c$ flavors reproduce the known physics.

The nonlinearities introduced by the baryonic operators make the computation
of the matrix model partition function challenging. Fortunately, when the
tree level superpotential can be expressed in terms of mesons, the matrix model
and gauge theory can be related more directly \cite{JADE}. Thus, adding in the matrix
path integral a constraint which identifies the matrix model quark bilinears with the
gauge theory mesons allows one to compute directly the gauge theory superpotential
with the corresponding mesons integrated in. This proposal was proved using the
geometric construction of the matrix model \cite{rtw}, using the symmetries of
the gauge theory \cite{brr}, and by explicitly integrating in quarks \cite{FEN5}.

Thus, the free energy which gives the superpotential of a theory with both massive and
massless flavors is given by
\be
e^{\F}=\int DQ~ D{\tilde Q} \delta(Q_{\hat \imath}^a{\tilde Q}_a^{\hat \jmath} -
M_{\hat \imath}^{\hat \jmath}) e^{-W_{\rm tree}(Q_{\hat \imath}, Q_\alpha)}
\Big|_{\rm planar+1\,boundary}
\ee
where $Q_\alpha $ are the massive quarks and $Q_{\hat \imath}$ are the
massless ones.

This expression also allows us to compute the free energy in the presence of baryons,
as long as $N_f=N_c+1$. Indeed, choosing two massive quarks, it is possible to
sum up all Feynman diagrams involving them and obtain a result which only depends
on quark bilinears. \footnote{This little miracle happens only for $N_f=N_c$
or $N_f=N_c+1$. If $N_f\ge N_c+2$  one needs to choose more than two massive
flavors, which makes the matrix integral non-Gaussian and rather hard to
compute.} At this stage the $\delta$-function
constraint can be easily enforced and we are left with computing the integral of the
constraint.

If the dimension of the matrices $M_c$ is larger than $N_f$ (which is always the case in large $M_c$ limit) this integral is \cite{JADE}:
\be
\int DQ_{\hat \imath} D{\tilde Q}^{\hat \jmath}
\delta({Q_{\hat \imath}^a{\tilde Q}_a^{\hat \jmath}-
M_{\hat \imath}^{\hat \jmath}}) =
e^{M_c\ln (\det {\hat M}/\Lambda^{ 2N_f})
-N_f\ln (\det {\hat M}/\Lambda^{ 2N_f}) -N_f M_c\ln M_c},
\ee
where $\Lambda$ is a cutoff. This result however contains both leading and subleading terms in $N_f/M_c$. In particular the
logarithm in the second term in the exponent is proportional to the number of flavors, ${N_f}$ (the determinant
is taken over flavor indices, and therefore is of order $M^{N_f}$), and therefore this term is of order $N_f^2$. Hence, it is generated by a multi-boundary diagram with insufficient gauge index loops, and should not be included in the free energy.

Identifying the matrix model 't~Hooft coupling with the gauge theory glueball
superfield and taking into account the clarifications above, the contribution of the planar and 1-boundary
diagrams to the above integral becomes:
\be
\int DQ_{\hat \imath} D{\tilde Q}^{\hat \jmath}
\delta({Q_{\hat \imath}^a{\tilde Q}_a^{\hat \jmath}-
M_{\hat \imath}^{\hat \jmath}}) =
e^{S\ln (\det {\hat M}/\Lambda^{ 2N_f}) - N_f S \ln (S/\Lambda^3)}
\label{delta}
\ee
This equation will be one of the important ingredients in our comparison
of matrix model and gauge theory results.


\section{Integrating-Out All Flavors\label{all_out}}

In this section we compute the gauge theory effective superpotential
at its critical points. The unbroken symmetries determine its value up to an unknown function of one variable. By requiring consistency with
the high energy theory, we
construct a differential equation for this function. We first solve
it for the special case of an $U(2)$ theory with three flavors and then turn to the general
case.

\subsection{Symmetries and Consistency Constraints\label{symmetries}}

We start with a tree level superpotential with
mass terms and baryon source terms for all flavors:
\begin{equation}
  W_{\it tree} = m^i_{j} Q_i \tilde{Q}^{{j}}
  + b_i B^i + \tilde{b}^{{j}} \tilde{B}_{{j}}.
\label{w-tree}
\end{equation}

The total superpotential is the sum of this tree level superpotential and of the dynamically generated superpotential
\be
W_{\it dyn}=-{1\over \Lambda^{2N_c-1}}\left(\det M-BM{\tilde B}\right).
\label{w-dyn}
\ee
The quantum numbers of the sources are

\begin{center}
\begin{tabular}{|c|cccccc|}
\hline
 & $U(1)_R$ & $SU(N_f)_Q$ & $SU(N_f)_{\tilde{Q}}$ & $U(1)_B$ & $U(1)_A$ & $D$
\\ \hline
$Q$ & ${1\over N_f}$ & $N_f$ & 1 & $+1$ & $+1$ & 1\\
$\tilde{Q}$ & ${1\over N_f}$ & 1 & $N_f$ & $-1$ & $+1$ & 1\\
$M$ & ${2\over N_f}$ & $N_f$ & $N_f$ & 0 & $+2$ & 2\\
$B$ & $1-{1\over N_f}$ & $\overline{N_f}$ & 1 & $N_f-1$ & $N_f-1$ & $N_f-1$\\
$\tilde{B}$ & $1-{1\over N_f}$ & 1 & $\overline{N_f}$ & $-N_f+1$ & $N_f-1$ & $N_f-1$
\\\hline
$m$ & $2-{2\over N_f}$ & $\overline{N_f}$ & $\overline{N_f}$ & 0 & $-2$ & 1\\
$b$ & $1+{1\over N_f}$ & $N_f$ & 1 & $-N_f+1$ & $-N_f+1$ & $-N_f+4$\\
$\tilde{b}$ & $1+{1\over N_f}$ & 1 & $N_f$ & $N_f-1$ & $-N_f+1$ & $-N_f+4$\\
$\Lambda^{2N_f-3}$ & 0 & 1 & 1 & 0 & $2N_f$ & $2N_f-3$\\
\hline
\end{tabular}
\end{center}
Using these quantum numbers, we can determine the form of the allowed
superpotential terms after integrating out all flavors.

All superpotential terms are functions of
$\Lambda^{2N_f-3}$, $b$, $\tilde{b}$, and $m$.  To construct
invariants under the non-abelian flavor symmetries, the only allowed
building blocks are $b_i m^i_{j} \tilde{b}^{{j}}$ and $\det m$.
The $U(1)_R$ invariant combination of these is
$(b_im^i_j\tilde{b}^j)^{N_f-1} / (\det m)^2$.  Its $U(1)_A$ charge is
$(-2N_f)(N_f-1) - 2(-2)N_f = -2N_f^2+6N_f$.  Therefore
the combination
\begin{equation}
  \frac{(b_i m^i_{j} \tilde{b}^{{j}})^{N_f-1}}
  {(\det m)^2} (\Lambda^{2N_f-3})^{N_f-3}
\end{equation}
is invariant under all symmetries, and is dimensionless as well.

The existence of the gluino condensate implies that in the absence of baryonic
sources, $b=\tilde{b}=0$, the superpotential is 
$(N_f-1)[({\det}m)\Lambda^{2N_f-3}]^{1/(N_f-1)}$.  Therefore the possible
form of the superpotential in the presence of baryon source terms is
\begin{equation}
  W_{\it eff} = (N_f-1) [(\det m)\Lambda^{2N_c-1}]^{1/N_c}
  f\left(\frac{(b_i m^i_{j} \tilde{b}^{{j}})^{N_f-1}}
  {(\det m)^2} (\Lambda^{2N_f-3})^{N_f-3}\right)~~,
\label{p1}
\end{equation}
where $f(x)$ is a function we want to determine.

In the limit of infinite mass parameter $m$ this theory reduces to a pure $N=1$ gauge
theory. In this case we know that there are $N_c$ vacua and the values of the
superpotential at the critical points differ by roots of unity of order $N_c$. In this limit,
the argument of the function $f$ in the equation above vanishes, while its coefficient
can be identified with the dynamical scale of the resulting theory. Thus, in order to
recover the expected gauge theory results, we must impose the boundary condition
\be
f(0)=\omega_{N_c}^k,\,k=0,\dots,N_c-1~~~~~ {\rm with}~~~~
\omega^{N_c}=1~~.
\label{bc}
\ee

%
%

The expectation values of the moduli are obtained by differentiating this effective
superpotential with respect to the
sources. \footnote{We use the simplified notation
$(b_i m^i_{j} \tilde{b}^{{j}})\equiv(b m \tilde{b})$}
\begin{eqnarray}
    M_{i}^{j} &=& \frac{\partial W_{\it eff}}{\partial m^i_{j}}
%
%
  =
  [(\det m)\Lambda^{2N_f-3}]^{1/(N_f-1)} \times\nonumber \\
  & &
  \Biggl( (m^{-1})^{i{j}} f(x)
  + (N_f-1)
  x f'(x) \left(
    \frac{(N_f-1)b_i \tilde{b}^{{j}}}{(b m \tilde{b})}
    - 2 (m^{-1})^{i{j}} \right) \Biggr) \\
    B^{i} &=& \frac{\partial W_{\it eff}}{\partial b_i}
  = (N_f-1)^2 [(\det m)\Lambda^{2N_f-3}]^{1/(N_f-1)}
  x f'(x)
  \frac{m^i_{j} \tilde{b}^{{j}}}
  {(b m \tilde{b})}\\
    \tilde{B}_{{j}} &=& \frac{\partial W_{\it eff}}
    {\partial \tilde{b}^{{j}}}
  = (N_f-1)^2 [(\det m)\Lambda^{2N_f-3}]^{1/(N_f-1)}
  x f'(x) \frac{b_i m^i_{j}} {(b m \tilde{b})}
\end{eqnarray}
Therefore, we find
\begin{eqnarray}
  \lefteqn{ B^i M_{i}^{j} =
    (N_f-1)^2 [(\det m)\Lambda^{2N_f-3}]^{2/(N_f-1)} x f'(x)\times}
  \nonumber \\
  & &
  \Biggl( (m^{-1})_{k}^{{j}} f(x)
  + (N_f-1)
  x f'(x) \left(
    \frac{(N_f-1)b_k \tilde{b}^{{j}}}{(b m \tilde{b})}
    - 2 (m^{-1})_{k}^{j} \right) \Biggr)
  \frac{m^{k}_{l} \tilde{b}^{{l}}}
  {(b m \tilde{b})} \nonumber \\
  &=& (N_f-1)^2 [(\det m)\Lambda^{2N_f-3}]^{2/(N_f-1)} x f'(x)
  \frac{\tilde{b}^{{j}}}{(b m \tilde{b})}\times \nonumber \\
  & & ~~~~~~( f(x) + (N_f-1) (N_f-3) x f'(x) ).
\end{eqnarray}

One of the equations of motion derived from $W_{\it tree}+ W_{\it dyn}$ (\ref{w-tree},
\ref{w-dyn}) imposes the following relation:
\begin{equation}
  B^i M_{i}^{j} = - \Lambda^{2N_f-3} \tilde{b}^{{j}}~~.
\end{equation}
Therefore,
\begin{equation}
  (N_f-1)^2 x^{-1/(N_f-1)} x f'(x)
  \left[ f(x) + (N_f-1) (N_f-3) x f'(x) \right]
  = - 1.
\label{geneq_all}
\end{equation}

This equation is special for $N_f=3$, as the secont term vanishes.
We begin in the following subsection by analyzing this case, and defer the general
discussion to section 8. We then give the matrix model description of the $SU(2)$
theory and compare the results.

\subsection{$SU(2)$ with three flavors:~Field Theory \label{su2_ft}}

For $N_f=3$ the superpotential (\ref{p1}) is
\begin{equation}
  W = 2 [(\det m)\Lambda^{3}]^{1/2}
  f\left(\frac{(b_i m^i_{j} \tilde{b}^{{j}})^{2}}
  {(\det m)^2}\right).
\end{equation}

Let us notice that the argument of $f(x)$ does not depend on $\Lambda$. This
is easy to understand. Because of the Lie algebra identification $SU(2)\simeq Sp(1)$,
the baryons can be interpreted as mesons in the $Sp(1)$ theory.  The mass matrix is
\begin{equation}
  \left(
    \begin{array}{cc}
      \epsilon^{ijk}b_k & m^i_j \\
      -m_{j}^{i} & \epsilon_{ijk} \tilde{b}^{{k}}
    \end{array} \right)
\end{equation}
and its Pfaffian can be perturbatively expanded around $\det m$ in
inverse powers of $m$.  The function $f(x)$ must be precisely this
expansion.  Therefore it is a polynomial of $\sqrt{x}$.

Indeed, equation (\ref{geneq_all}) reduces to
\begin{equation}
4 x^{1/2} f'(x) f(x)
  = - 1.
\end{equation}
which implies that $f(x)$ is given by
\begin{equation}
  f(x) = \pm (C-x^{1/2})^{1/2},
\end{equation}
The integration constant $C$ is fixed to unity by the boundary conditions (\ref{bc}).
Then, the effective superpotential becomes
\begin{equation}
  W = \pm 2 [(\det m)\Lambda^{3}]^{1/2}
  \left(1 - \frac{b_i m^i_{j} \tilde{b}^{{j}}}
    {\det m}\right)^{1/2}
  = \pm 2 \left[ (\det m) - (bm{\tilde b})
  \right]^{1/2} \Lambda^{3/2}~~.
\label{u2ft}
\end{equation}
The combination in the square bracket is precisely the Pfaffian of the
mass matrix including the baryon source terms.
In the next section we will recover this result from matrix model computations.


\section{$SU(2)$ with three flavors; Matrix Model \label{mm2}}


In this section we describe $SU(2)$ supersymmetric QCD with 3 flavors using the matrix model. Since the baryon
operators in this theory are  bilinear in quarks,
the matrix model free energy can be computed directly. We will find
that, after integrating out the glueball superfield, the effective superpotential
agrees with the field theory result given in Eq. (\ref{u2ft}). \footnote{For larger $N_c$ the
matrix model is sufficiently complicated to render challenging
the direct recovery of the field theory results. We will return to these questions
in section \ref{general_all}.}

As briefly stated in the previous section, for an $SU(2)$ theory with three flavors
the tree level superpotential is
\be
W_{\it tree}=m_i^jQ_i^a{\tilde Q}_a^j+b_i\epsilon^{ijk}\epsilon_{ab}Q_j^aQ_k^b
+{\tilde b}^i\epsilon_{ijk}\epsilon^{ab}{\tilde Q}^j_a{\tilde Q}^k_b~~.
\ee
To compute the partition function it is useful to rewrite this expression as
\be
W_{\it tree}={1\over 2}{\cal Q}^T K_{U(2)}{\cal Q}
\ee
where
\be
K_{U(2)}=\pmatrix{b_i\epsilon^{ijk}\otimes \epsilon_{ab} &
m_j^k \otimes \delta_a^b\cr
m_j^k \otimes \delta_a^b & {\tilde b}^i
\epsilon_{ijk}\otimes \epsilon^{ab} \cr}~~~~{\rm and}~~~~
{\cal Q}=\pmatrix{Q_k^b\cr {\tilde Q}_b^j}
\ee

The 1-boundary free energy is given by the logarithm of the determinant of $K_{U(2)}$.
This can be easily computed and it gives:
\be
\det K_{U(2)}=\left(\det m-(b m {\tilde b})\right)^{2\times 2}
\ee
where in the exponent the first factor of $2$ is due to the fact that we integrated over
two types of fields, $Q$ and ${\tilde Q}$, while the second factor of $2$ represents the
number of colors.

In principle one should worry about isolating the planar diagram contribution to the free energy. 
Fortunately, for $SU(2)$, all the
diagrams are planar. \footnote{This is due to the fact that both
$\delta_a^b$ as well as $\epsilon_{ab}$ are invariant tensors. The non-planarity can
in principle arise due to insertions of a baryonic operator in the Feynman diagram,
but the antisymmetry of $\epsilon_{ab}$ can be used to transform it into a planar one.}

Combining this with the Veneziano-Yankielowicz term yields the effective superpotential:
\bea
W_{\it eff}=N_cS\left(1-\ln {S\over \Lambda^3}\right)
-S\ln{1\over \Lambda^3}\left(\det m-(b m{\tilde b})\right)
\eea

To compare with the field theory result we must integrate out $S$. This gives
\be
W_{\it eff}=\pm 2\left(\det m-(b m {\tilde b})\right)^{1/2}\Lambda^{3/2}
\label{finegsu2_1}
\ee
which precisely matches the field theory result.

Perhaps this agreement should not appear surprising, since for a $U(2)$ gauge
group the mesons and baryons have similar structure. However, the computations
which led to the two results are substantially different; this seems to imply that the
agreement is somewhat nontrivial. Another point worth emphasizing is that
{\em all} matrix model diagrams contributed to the effective superpotential. The origin of
this fairly surprising fact is again the bilinearity of the baryons. This will not
happen in the general case to which we return in section \ref{mm_gen_2}.


\section{Integrating-Out Two Flavors - The Elegant Way \label{slick}}

As discussed in section \ref{generalities}, our goal is to match the gauge theory
effective superpotential after integrating out two quarks with
the matrix model predictions.
Let us therefore begin with the appropriate computation on the gauge theory side.
There are two ways to achieve our goal. In this section, using symmetry arguments,
we constrain the form of the effective potential after integrating out two quarks
and then derive certain constraints on the unknown functions. Solving these
constraints leads to our result. In the next section we rederive the same result by directly
integrating out the appropriate fields.

Since we only give mass to two of the flavors, the tree level superpotential is
\begin{equation}
  W = m^{\alpha}_{\beta} Q_{\alpha} \tilde{Q}^{{\beta}}
  + b_i B^i + \tilde{b}^{{j}} \tilde{B}_{{j}}~~,
\end{equation}
where
$\alpha,\beta=N_c,N_c+1$. Hereafter we distinguish the indices of the massive and massless flavors:
$\alpha,\beta=N_c,N_c+1$, and ${\hat \imath}, {\hat \jmath}=1, \cdots, N_c-1$ respectively.

The superpotential has a tree level part
\begin{equation}
  W = m^{\alpha}_{{\beta}} M_{\alpha}^{{\beta}}
  + b_\alpha B^\alpha + b_i B^i
  + \tilde{b}^{{\beta}} \tilde{B_{{\beta}}}
  + \tilde{b}^{{j}} \tilde{B}_{{j}}~~,
\label{wtree}
\end{equation}
and a non-perturbatively generated part
\begin{equation}
  \frac{1}{\Lambda^{2N_c-1}}
  \left( B^{\hat \imath} M_{\hat\imath}^{\hat\jmath} \tilde{B}_{\hat\jmath}
    + B^\alpha M_\alpha^{\hat \jmath} \tilde{B}_{\hat \jmath}
    + B^{\hat \imath} M_{\hat \imath}^{\beta} \tilde{B}_{ \beta}
    + B^\alpha M_{\alpha}^{\beta} \tilde{B}_{\beta}
    - \det M
    \right)~~.
\label{wdyn}
\end{equation}

\subsection{Preliminaries}

Our goal is to find the effective superpotential after integrating
out the two massive flavors.  This superpotential is a
function of $b_\alpha$, $b_{\hat\imath}$, $\tilde{b}^{{\beta}}$,
$\tilde{b}^{\hat{\jmath}}$, $m$ and $M_{\hat\imath}^{\hat{\jmath}}$.  In the absence of baryonic source terms, it is easy to find the solution
\begin{eqnarray}
  & & B^\alpha = B^{\hat\imath} = \tilde{B}_{{\beta}} = \tilde{B}_{{\hat\jmath}} = 0~~,
  \\
  & & M^{\hat\imath}_{{\beta}} = M^{\alpha}_{\hat {\jmath}} = 0~~, \\
  & & M^{\alpha}_{{\beta}}
  = \frac{(m^{-1})^{\alpha}_{{\beta}} (\det   m)\Lambda^{2N_c-1}}
  {\det \hat{M}}~~,
\end{eqnarray}
where, $(m^{-1})$ is the inverse of the two-by-two mass matrix, and
$\hat{M}$ is the meson matrix constructed out of the remaining flavors.
The resulting effective superpotential is
\begin{equation}
  W = \frac{(\det   m)\Lambda^{2N_c-1}}{\det \hat{M}}~~.
\end{equation}
which is the expected Affleck--Dine--Seiberg
superpotential.

In the general case the quantum numbers under the
$SU(N_f-2)_Q \times SU(2)_Q \times U(1)_Q \subset SU(N_f)$ flavor
symmetry and its counterpart for $\tilde{Q}$, force the
superpotential to take the form
\begin{equation}
  W = \frac{(\det   m)\Lambda^{2N_c-1}}{\det \hat{M}}
  f(x,y), \label{fxy}
\end{equation}
where the invariants $x$ and $y$ are
\begin{eqnarray}
  x &=& \frac{(bm\tilde{b})(\det \hat{M})^2}
  {(\det m)^2\Lambda^{2N_c-1}}~~, \\
  y &=& \frac{(b\hat{M}^{-1}\tilde{b})\det \hat{M}}{(\det m)}~~.
\end{eqnarray}
with $(bm\tilde{b}) \equiv b_\alpha
m^{\alpha}_{{\beta}} \tilde{b}^{{\beta}}$ and
$(b\hat{M}^{-1}\tilde{b}) \equiv b_{\hat \imath}
(\hat{M}^{-1})^{{\hat \imath}}_{{\hat \jmath}} \tilde{b}^{\hat \jmath}$.  We
require that the superpotential be regular in the limit of no baryon
sources and also at weak coupling $\Lambda \rightarrow 0$. Therefore, the function $f(x,y)$ can be at most linear in $x$, and
hence
\begin{equation}
  W = \frac{(\det   m)\Lambda^{2N_c-1}}{\det \hat{M}} g(y)
  + \frac{(bm\tilde{b})\det \hat{M}}{\det m} h(y)~~.
\label{undetgh}
\end{equation}

In order to obtain the explicit forms of $g(y)$ and $h(y)$ it is useful to 
consider several limiting cases.

\subsection{$b_{\hat \imath} = \tilde{b}^{{\hat \jmath}} = 0$ with
${\hat \imath},{\hat \jmath}=1,\cdots,N_c-1$}

In this case, the $SU(N_f-2)_Q \times SU(N_f-2)_{\tilde{Q}}$
symmetry is unbroken.  Hence,
\begin{eqnarray}
  & & M^{{\hat\imath}}_{{\beta}} = M^{\alpha}_{\hat \jmath} = 0, \\
  & & B^{\hat \imath} = \tilde{B}_{{\hat \jmath}} = 0.
\end{eqnarray}
The equations of motion are
\begin{eqnarray}
  \label{eq:37}
  && B^{\alpha} \tilde{B}_{{\beta}} - (M^{-1})^{\alpha}_{{\beta}}
  (\det M) + \Lambda^{2N_c-1} m^{\alpha}_{{\beta}} = 0,\\
  && B^{\alpha} = (M^{-1}){}^{\alpha}_{{\beta}} \tilde{b}^{{\beta}},\\
  && \tilde{B}_{{\beta}} = b_{\alpha} (M^{-1})_{\beta}^{{\alpha}}.
\end{eqnarray}
where $(M^{-1})_{\alpha\tilde{\beta}}$ is defined only in the
two-by-two block.  Substituting the solutions from the last two
equations into the first one, we find
\begin{equation}
  [(M^{-1})^{\alpha}_{{\beta}} \tilde{b}^{{\beta}} ]
  [b_{\gamma} (M^{-1})^{\gamma}_{{\beta}}] - (M^{-1})^{\alpha}_{{\beta}}
  (\det M) + \Lambda^{2N_c-1} m^{\alpha}_{{\beta}} = 0.
\end{equation}
This is an equation for two-by-two matrices and hence there are four
unknowns.  On symmetry grounds we take the following ansatz:
\begin{equation}
  (M^{-1})^{\alpha}_{{\beta}} = \alpha m^{\alpha}_{{\beta}}
  + \beta (m\tilde{b})^\alpha (bm)_{{\beta}}
\end{equation}
where $\alpha$, $\beta$ are function of the invariants.  Apparently this system
is over-constrained, as there are four equations for two unknowns.  However, a
solution exists and is given by
\begin{eqnarray}
  \alpha &=&
  \frac{-(\det M)(\det m)}
  {- (bm\tilde{b}) (\det M)^2
    + (\det m)^2 \Lambda^{2N_c-1} }\, , \\
  \beta &=&
  \frac{-(\det M)^3}
  {(\det m) ( (bm\tilde{b})(\det M)^2
    - (\det m)^2 \Lambda^{2N_c-1}) \Lambda^{3 - 2N_f}} \, .
\end{eqnarray}
Using this solution, the superpotential is given by
\begin{equation}
  \label{eq:42}
  W_{\it eff} =\frac{( \det m ) {\Lambda }^{2N_c-1}  }
  {(\det \hat{M})}
  - \frac{(\det \hat{M}) ( bm\tilde{b}) }
  {( \det m )}~~,
\end{equation}
which is precisely what we expected from the symmetry considerations,
except that we now determined the coefficient $-1$ for the second
term.  This determines the boundary condition $h(0)=-1$.

\subsection{
$b_\alpha=\tilde{b}^{{\beta}} = 0$ with $\alpha,\beta=N_c,N_c+1$}

The next simple case is $b_\alpha=\tilde{b}^{{\beta}} = 0$, when the only parameters in the effective
superpotential are $m_{\alpha}^{{\beta}}$, $b_i$,
$\tilde{b}^{j}$, $M_{{\hat \imath}}^{{\hat \jmath}}$. Hence there are no doublet breaking parameters 
of $SU(2)_Q \times SU(2)_{\tilde{Q}}$.
This immediately gives
\begin{equation}
  M_{{\hat \imath}}^{{\beta}} = M^{\alpha}_{{\hat \jmath}} = B^\alpha =
  \tilde{B}_{{\beta}} = 0.
\end{equation}
The equations of motion can be easily solved,
\begin{eqnarray}
  \label{eq:32}
  & & \tilde{B}_{{\hat \jmath}}
  = - \Lambda^{2N_c-1} b_{\hat \imath} (\hat{M}^{-1})^{{\hat \imath}}_{{\hat \jmath}}, \\
  & & B_{{\hat \imath}}
  = - \Lambda^{2N_c-1} (\hat{M}^{-1})_{{\hat \imath}{\hat \jmath}}
\tilde{b}^{{\hat \jmath}}, \\
  & & M^{\alpha}_{{\beta}}
  = (m^{-1})^{\alpha}_{{\beta}}
  \frac{(\det m) \Lambda^{2N_c-1}}{(\det \hat{M})},
\end{eqnarray}
where $\hat{M}$ is the meson matrix for the remaining
$N_f-1$ flavors.  Substituting the solutions to the superpotential, we
find
\begin{equation}
  \label{eq:33}
  W_{\it eff} = \frac{(\det m) \Lambda^{2N_c-1}}{\det \hat{M}}
  - \Lambda^{2N_c-1} b_i (\hat{M}^{-1})^{{\hat \imath}}_{{\hat \jmath}}
\tilde{b}^{{\hat \jmath}}~~,
\end{equation}
and therefore $g(y)=1-y$. The only remaining function to be determined is $h(y)$.

\subsection{General Case}

Putting together what we have learned so far, the superpotential is
\begin{equation}
  W_{\it eff} = \frac{(\det m) \Lambda^{2N_c-1}}{\det \hat{M}}
  - \Lambda^{2N_c-1} (b{\hat M}^{-1}{\tilde b})
  + \frac{(bm\tilde{b})\det \hat{M}}{\det m} h(y),
\end{equation}
with $h(0)=-1$ and
\begin{equation}
  y = \frac{(b\hat{M}^{-1}\tilde{b})\det \hat{M}}{(\det m)}.
\end{equation}
From this superpotential we can obtain the vacuum expectation values of the $ M_{\alpha}^{{\beta}}$ mesons and of the baryons:
\begin{eqnarray}
  \label{eq:34}
  M_{\alpha}^{{\beta}} &=&
  \frac{\partial W_{\it eff}}{\partial m^{\alpha}_{{\beta}}}
  = \frac{(m^{-1})_{\alpha}^{{\beta}} (\det m)
    \Lambda^{2N_c-1}}{\det \hat{M}}
  + \frac{(b_\alpha \tilde{b}^{{\beta}}) \det \hat{M}}{{\det}m} h(y) \nonumber \\
  & &
  - (m^{-1})_{\alpha}^{{\beta}} \frac{(bm\tilde{b}){\det}\hat{M}}{\det m} h(y)
  - \frac{(bm\tilde{b})\det \hat{M}}{\det m} yh'(y)
  (m^{-1})_{\alpha}^{{\beta}},\\
  \tilde{B}_{{\hat \jmath}} &=& \frac{\partial W_{\it eff}}{\partial
    \tilde{b}_{{\hat \jmath}}}
  = -\Lambda^{2N_c-1} (b\hat{M}^{-1})_{{\hat \jmath}}
  + \frac{(bm\tilde{b})\det \hat{M}}{\det m} yh'(y)
  \frac{(b\hat{M}^{-1})_{{\hat \jmath}}}{(b\hat{M}^{-1}\tilde{b})},\\
  \tilde{B}_{{\beta}} &=& \frac{\partial W_{\it eff}}{\partial
    \tilde{b}^{{\beta}}}
  = - \frac{(bm)_{{\beta}} \det \hat{M}}{\det m} h(y),
\end{eqnarray}
The one piece of information we cannot
obtain from this superpotential is the vacuum expectation value of the 
off-diagonal mesons. By symmetry considerations it must be of the form
\begin{equation}
  \label{eq:35}
  M_{\alpha}^{ {\hat \jmath}} = \alpha(x,y) b_\alpha \tilde{b}^{{\hat \jmath}},
\end{equation}
and similarly for $M_{{\hat \imath}}^{{\beta}}$. To determine the unknown 
functions $\alpha(x,y)$ and $h(y)$ one must use the equations of motion. We start with
\begin{eqnarray}
  \label{eq:36}
  0 &=& \frac{\partial W}{\partial B^\alpha}
  = \frac{M_{\alpha}^{{\beta}} \tilde{B}_{{\beta}}
    + M_{\alpha}^{{\hat \jmath}} \tilde{B}_{{\hat \jmath}}}{\Lambda^{2N_c-1}}
  + b_\alpha \nonumber \\
  &=& \frac{1}{\Lambda^{2N_c-1}} \Biggl\{
    b_\alpha \Lambda^{2N_c-1} h(y)
    + b_\alpha \frac{(bm\tilde{b}) (\det \hat{M})^2}{(\det m)^2}
    y h'(y) h(y) \nonumber \\
    & &
    + \alpha \left[ -\Lambda^{2N_c-1} (b\hat{M}^{-1}\tilde{b}) b_\alpha
      + \frac{(bm\tilde{b})\det \hat{M}}{\det m} y h'(y) b_\alpha
      \right] \Biggr\}
    + b_\alpha.
\end{eqnarray}
This leads to the differential equation
\begin{equation}
  \label{eq:38}
  1 + h(y) + x y h'(y) h(y) + \alpha \frac{\det m}{\det \hat{M}}
  [-y+xyh'(y)] = 0.
\end{equation}
Another useful equation is
\begin{eqnarray}
  \label{eq:39}
  0 &=& \frac{\partial W}{\partial B^i}
  = \frac{M_{i}^{{\beta}} \tilde{B}_{{\beta}}
    + M_{i}^{{\hat \jmath}} \tilde{B}_{{\hat \jmath}}}{\Lambda^{2N_c-1}} + b_i
  \nonumber \\
  &=& \frac{b_i}{\Lambda^{2N_c-1}} \left\{
    \alpha  \frac{(bm\tilde{b}) \det \hat{M}}{\det m} h(y)
    - \Lambda^{2N_c-1} \right.\nonumber \\
  & & \left.~~~~~~~~~~~~
       + \frac{(bm\tilde{b}) \det \hat{M}}{\det m \,(b\hat{M}^{-1}\tilde{b})} y h'(y)
       \right\}   + b_i.
\end{eqnarray}
This leads to another differential equation
\begin{equation}
  \label{eq:40}
  \alpha \frac{\det m}{\det \hat{M}} h(y) + h'(y) = 0.
\end{equation}

Solving for $\alpha$ from the second equation and substituting it into
the first one, we obtain
\begin{equation}
  \label{eq:41}
  1 + h(y) - x y h'(y) h(y) + (y-xyh'(y)) \frac{h'(y)}{h(y)} = 0.
\end{equation}
Because this equation has to hold for any $x$, it gives two
equations for $h(y)$,
\begin{eqnarray}
  \label{eq:43}
  & & 1 + h(y) + y \frac{h'(y)}{h(y)} = 0, \\
  & & - y h'(y) h(y) + y h'(y) \frac{h'(y)}{h(y)} = 0.
\end{eqnarray}
It is non-trivial that two different non-linear differential equations
have a consistent solution.  The first equation gives
\begin{equation}
  \label{eq:44}
  \frac{dh}{h^2+h} = - \frac{dy}{y},
\end{equation}
and hence
\begin{equation}
  \label{eq:45}
  \log \left| \frac{1+h(y)}{h(y)} \right|
  = \log |y| + {\rm const}.
\end{equation}
Together with the boundary condition $h(0)=-1$, this leads to the
solution
\begin{equation}
  \label{eq:46}
  h(y) = - \frac{1}{1-y}.
\end{equation}
On the other hand, the second equation gives
\begin{equation}
  \label{eq:47}
  \frac{dh}{h^2} = -dy,
\end{equation}
and hence
\begin{equation}
  \label{eq:48}
  h(y) = -\frac{1}{1-y}.
\end{equation}
Both equations give the same solution, which confirms our result.
\footnote{We can also determine $\alpha(x,y)$:
\begin{equation}
  \label{eq:49}
  \alpha (x,y) \frac{\det m}{\det \hat{M}} = - \frac{1}{1-y},
\end{equation}
and hence
\begin{equation}
  \label{eq:50}
  M_{\alpha}^{{\hat \jmath}} = - \frac{b_\alpha \tilde{b}^{{\hat \jmath}}}{1-y}
  \frac{\det \hat{M}}{\det m}
~~~;~~~~~
M_{\hat \jmath}^{{\alpha}} = - \frac{b_{\hat \jmath} \tilde{b}^{{\alpha}}}{1-y}
  \frac{\det \hat{M}}{\det m}
~~.
\end{equation}
These expressions are necessary for integrating the two flavors back in.}
Therefore, the effective superpotential after integrating out two quarks is
\bea
W_{\it eff}&=&{\Lambda^{2N_c-1}}\left[
{ \det m\over \det {\hat M}}-(b{\hat M}^{-1}{\tilde b})\right]
-
{(b m {\tilde b})\det {\hat M}
\over \det m-(b{\hat M}^{-1}{\tilde b})\det{\hat M}}.
\label{ft}
\eea


\section{Integrating-Out Flavors - The Laborious Way\label{hard}}

In this section we will recover the results of the previous section using a
different method: instead of using symmetries to constrain the final form of
the effective superpotential, we will just directly solve the classical equations
of  motion and then evaluate the initial superpotential at these values of the fields.
Starting from (\ref{wtree}, \ref{wdyn})
\bea
\label{W}
W_{\it eff}\!\!&= &\!\!M_{\alpha }^{ {\beta}} m_{ {\beta}}^{ \alpha} +
b_{\hat \imath} B^{\hat \imath} + \ti b^{ {\hat \imath}} \ti B_{ {\hat \imath}} +
b_{\alpha} B^{\alpha} + \ti b^{ \alpha} \ti B_{ \alpha}\\
&+&\!\!
{1\over \Lambda^{2N_c-1} }\left[B^{{\hat \imath}} M_{{\hat \imath} }^{ {\hat \jmath}}
\ti B_{ {\hat \jmath}} +
B^{{\hat \imath}} M_{{\hat \imath} }^{ \alpha} \ti B_{ \alpha} +
B^{\alpha} M_{\alpha }^{
{\hat \jmath}} \ti B_{ {\hat \jmath}} +
B^{\alpha} M_{\alpha }^{ {\beta}} \ti B_{ {\beta}} -\det M\right]~~.\nonumber
\eea
the equations of motion are:
\bea
B_i \ti B^j - \bar M_i^j + m_i^j\Lambda^{2N_c-1}&=&0 \label{bb}\\
B_i M^i_j + \ti b_j\Lambda^{2N_c-1} &=&0 \label{bm}
\eea
where $i,j=1...N_c+1$, $\bar M_{i}^{j}$ is the cofactor, and only
$m_{\alpha}^{\beta} \neq 0$. We split the $(N_c+1) \times (N_c+1)$
matrix $M_{i}^j$ into a $2\times 2$ block $M_{\alpha}^{\beta}$, and a
$(N_c-1) \times (N_c-1)$ block $\hat M_{\hat \imath }^{\hat \jmath}$. The
off diagonal blocks are $M_{\alpha }^{\hat \jmath}$ and $M_{\hat \imath }^{ \alpha}$
respectively.

Multiplying (\ref{bb}) by $M_{k}^i$ and using the fact that
$M_{i }^{ j} \bar M_{j}^{ k} = \det M \delta^{i}_{k}$ we find after a
few straightforward steps:
\bea
M_{{\hat \imath}}^{ \alpha} m_{ \alpha}^{ {\beta}} &= & b_{\hat \imath} B^{{\beta}}  \\
M_{\alpha}^{ {\hat \imath}} m_{ {\beta}}^{ \alpha} &= & \ti b^{ {\hat \imath}} \ti B_{ {\beta}}  \\
M_{{\gamma} }^{ \alpha} m_{ \alpha }^{{\beta}} &= & b_{\gamma} B^{{\beta}} + \delta_{{\gamma}}^{ {\beta}} \det M \Lambda^{-(2N_c-1)}
\label{r1}\\
M_{{\gamma}}^{\alpha} m_{{\beta}}^{ {\gamma}} &= & \ti b^{ \alpha} \ti B_{ {\beta}} +\delta^{\alpha}_{ {\beta}} \det M \Lambda^{-(2N_c-1)}
\label{r2}
\eea

Equations (\ref{r1}) and (\ref{r2}) give $\ti B_{ {\beta}} \ti b^{ \alpha} m_{ \alpha}^{ {\gamma}} =  B^{ {\gamma}} b_{ \alpha} m^{ \alpha}_{ {\beta}}$, which implies
\bea
B^{\gamma} &=& B~ m^{{\gamma}}_{ \alpha } ~ \ti b^{ \alpha} \Lambda^{2(2N_c-1)}\\
\ti B_{ \alpha} &=& B~ m^{{\gamma}}_{ \alpha } ~b_{{\gamma}}\Lambda^{2(2N_c-1)}
~~,
\eea
where $B$ is a parameter.

We will first  express all the expectation values in terms of $B$, and then use some of
the remaining equations of motion to relate $B$ and $\det M$.
The mesons are given by:
\bea
M_{{\beta} }^{ \alpha} &=& b_{\beta} \ti b^{ \alpha} ~B \Lambda^{2(2N_c-1)}+ (m^{-1})_{{\beta} }^{ \alpha} ~ \det M \Lambda^{-(2N_c-1)}\\
M_{\alpha }^{ {\hat \imath}} &=& b_{\alpha} \ti b^{ {\hat \imath}} ~B \Lambda^{2(2N_c-1)}\\
M_{{\hat \imath} }^{ \alpha} &=& b_{\hat \imath} \ti b^{ \alpha} ~B \Lambda^{2(2N_c-1)}
\eea
Combining these equations with equation (\ref{bm}) one finds the baryons $B^{{\hat \imath}}$:
\bea
B^{\hat \imath} &=& - (\hat M^{-1})^{{\hat \imath}}_{\hat \jmath} \ti b^{ {\hat \jmath}} 
(1+ X B^2)\Lambda^{2N_c-1} \\
\ti B_{ {\hat \imath}} &=& - b_{\hat \jmath} (\hat M^{-1})_{{\hat \imath}}^{{\hat \jmath}} 
(1+ X B^2)\Lambda^{2N_c-1}
\eea
where
\be
X \equiv (b m \ti b)\Lambda^{3(2N_c-1)}.
\label{X}
\ee
Substituting everything back into $W_{\it eff}$ we find
\bea
W_{\it eff}&=& - \Lambda^{2Nc-1} (b {\hat M}^{-1}\tilde{b}) (B^2 X+1)^2 \nonumber\\
&+&
{1\over \Lambda^{2Nc-1} }\left[
(B^2 X+ 1) \det M + BX(B^2 X + 3)\right]~~.
\label{weff}
\eea
where as before $(b {\hat M}^{-1}\tilde{b})$ is a shorthand  for
$b_{\hat \imath} (\hat M^{-1})^{{\hat \imath} }_{ {\hat \jmath}}
\ti b^{{\hat \jmath}}$.

The next step in our evaluation is to find the relation between $\det M$ and $B$.
Using the block decomposition of the meson matrix we outlined in the beginning,
it is not hard to find that $\det M$ can be expressed as:
\be
\det M = (\det \hat M_{{\hat \imath} }^{ {\hat \jmath}})
\det (M_{\alpha }^{ {\beta}} - M_{\alpha }^{ {\hat \jmath}} (\hat M^{-1})_{\hat \jmath}^
{\hat \imath} M_{{\hat \imath} }^{ {\beta}})
\label{dm}
\ee
After expressing all its components in terms of $B$, one can easily compute the
determinant of the $2\times 2$ matrix to be:
\be
\det (M_{\alpha }^{ {\beta}} - M_{\alpha }^{ {\hat \jmath}}
(\hat M^{-1})_{\hat \jmath}^{\hat \imath} M_{{\hat \imath} }^{ {\beta}}) =
{\det M^2 + X \left[B-\Lambda^{2(2N_c-1)}B^2 (b {\hat M}^{-1}\tilde{b})\right]\det M
\over \Lambda^{2(2N_c-1)} \det m } ~~.
\ee
We should note that if $k$ flavors were integrated out, the numerator on the
right-hand-side of the equation above should be replaced with
$ \det M^k + X (B-B^2 b {\hat M}^{-1}\tilde{b}) \det M^{k-1}$.
Thus, the first equation relating B and $\det M$ is:
\be
(\det M)^2 ={(\det M) (\det m )\Lambda^{2(2N_c-1)} \over (\det \hat M
)} - (\det
M) X (B-\Lambda^{2(2N_c-1)}B^2 (b {\hat M}^{-1}\tilde{b}))
\label{det}
\ee
To find the other relations between $B$ and $\det M$ we use the equation of motion:
\be
B^{\alpha} B_{\beta} + m^{\alpha}_{\beta}\Lambda^{2N_c-1} =
{\partial \det M \over \partial M_{\alpha}^{\beta}} =
(\det \hat M_{{\hat \imath} }^{ {\hat \jmath}})
{\partial  \det \left[M_{\alpha }^{ {\beta}} - M_{\alpha }^{ {\hat \jmath}}
(\hat M^{-1})_{\hat \jmath}^{\hat \imath}\hat  M_{{\hat \imath} }^{ {\beta}}\right]
\over \partial M_{\alpha}^{\beta}}
\label{eom}
\ee
Multiplying this equation by $( M_{\alpha }^{ {\beta}}  -
{M_{\alpha }^{ {\hat \jmath}} (\hat M^{-1})_{\hat \jmath}^{\hat \imath}
M_{{\hat \imath} }^{ {\beta}} })$ and using the fact that
$M_{i }^{ j} { \partial \det M \over \partial M^{j}_{ k}} = \det M \delta_{i}^{k}$
we obtain after a few steps:
\be
\det M = - B X -{1 \over B} + \Lambda^{2(2N_c-1)}(B^2 X + 1) (b {\hat M}^{-1}\tilde{b})~~.
\label{e1}
\ee
Again this relation is independent of the number of flavors integrated out.
One can also evaluate by hand the cofactors in (\ref{eom}), sum them with
$m_{\alpha}^{\beta}$, and obtain
\be
\det M =  {\det m ~\Lambda^{2(2N_c-1)} \over 2 \det \hat M
}(2+ B^2 X) -{ X \over 2} (B-\Lambda^{2(2N_c-1)}B^2 (b {\hat M}^{-1}\tilde{b}))
\label{e2}
\ee
The equations (\ref{det}, \ref{e1}, \ref{e2}) have a unique solution
\bea
\det M &=&{\det m ~\Lambda^{2(2N_c-1)}\over
\det \hat M
}  -
X \left[ B-\Lambda^{2(2N_c-1)}B^2 (b {\hat M}^{-1}\tilde{b})\right]~~\nonumber\\
B &=&  { \Lambda^{-2(2N_c-1)} \det \hat M \over {(b {\hat M}^{-1}\tilde{b}) \det \hat M -
\det m
}}~~.
\eea
which gives
\bea
 W_{\rm eff}
&=&\Lambda^{2Nc-1}  \left[
{\det  m \over \det \hat M}  -
(b{\hat M}{}^{-1}\ti b) \right]-{ (b m\ti b)\det \hat M
\over  \det m - (b{\hat M}{}^{-1}\ti b)~ \det \hat M}
\label{ft2}
\eea
We have thus recovered the effective superpotential (\ref{ft}) constructed in section \ref{slick}.
We now turn to the matrix model analysis of the theory and recover the same results.


\section{$SU(N_c)$ with $N_c+1$ flavors; Matrix Model\label{mm_gen_2}}

The tree level superpotential of the theory under consideration was described
in section \ref{symmetries}. Since the goal is summing all diagrams
containing two flavor fields, it is useful to rewrite it in the following form:
\bea
W_{\it tree}&=& m_\alpha^\beta Q_{\beta}^a{\tilde Q}^{\alpha}_a
+\epsilon^{\alpha\beta} b_\alpha Q_{\beta}^a V_a+\epsilon_{\alpha\beta}
{\tilde b}^\alpha {\tilde Q}^{\beta}_a
{\tilde V}^a \nonumber\\
&+&b_{\hat \imath}\epsilon^{\alpha\beta}Q_{\alpha}^aQ_{\beta}^b
V^{\hat \imath}_{ab}
+{\tilde b}^{\hat \imath}\epsilon_{\alpha\beta}{\tilde Q}^{\alpha}_a{\tilde Q}^{\beta}_b
{\tilde V}_{\hat \imath}^{ab}
\nonumber
\eea
where $\alpha$ and $\beta$ take the values $N_c$ and $N_c+1$, and
\bea
&&{V}_a
=\epsilon^{N_c, N_c+1,{\hat \imath}_1,\dots, {\hat \imath}_{N_c-1}}
\epsilon_{a a_1\dots a_{N_c-1}}Q_{{\hat \imath}_1}^{a_1}\dots
Q_{{\hat \imath}_{N_c-1}}^{a_{N_c-1}}\nonumber\\
&&{\hat \imath}_1,\dots,{\hat \imath}_{N_c-1}=1,\dots, N_c-1~~~~a_1,\dots,
a_{N_c-1}=1,\dots N_c
\eea
and similarly for ${\tilde V}^a$. Also,
\bea
&&{V}^{\hat\imath}_{b_1b_2}=
\epsilon^{N_c, N_c+1,{\hat \imath},{\hat \imath}_1,\dots, {\hat \imath}_{N_c-2}}
\epsilon_{b_1b_2 a_1\dots a_{N_c-2}}Q_{{\hat \imath}_1}^{a_1}\dots
Q_{{\hat \imath}_{N_c-2}}^{a_{N_c-2}}\nonumber\\
&&{\hat \imath}_1,\dots,{\hat \imath}_{N_c-2}=1,\dots, N_c-1~,~~
b_1,~b_2,~a_1,\dots,
a_{N_c-1}=1\dots N_c
\eea
and similarly  ${\tilde V}_{\hat \imath}^{ab}$.

For latter convenience let us point out that:
\be
Q_{\hat \imath}^a{V}_a=0
~~~~~~~~
{\tilde Q}^{\hat \imath}_a{\tilde V}^a=0
\label{zero}
\ee
where ${\hat \imath}$ and ${\hat \jmath}$ take values {\it only} from $1$ to $N_c-1$.
There is no constraint of this sort for $Q_{\hat \jmath}^a{V}^{\hat \imath}_{ab}$, etc.
However, one can see that
\be
{V}_a{V}_b{\tilde V}^{ab}_{\hat \imath}=0
~~~~{\rm and}~~~~
{\tilde V}^a{\tilde V}^b{V}_{ab}^{\hat \imath}=0
~~~~
(\forall)~~{\hat \imath}=1,\dots,N_c-1~~.
\ee

To systematically compute the integral it is useful to write the tree level superpotential
as a quadratic form.
This is easily done by introducing
\be
{\cal Q}=\pmatrix{
Q^{a}_{\alpha}\cr
{\tilde Q}_a^{\beta}\cr}
~~~~~~~~
{\cal V}=\pmatrix{\sum_{\gamma=N_c}^{N_c+1}
\epsilon^{\alpha\gamma}b_{\gamma} V_a \cr
\sum_{\gamma=N_c}^{N_c+1}
\epsilon_{\beta\gamma}{\tilde b}^{\gamma} {\tilde V}_a \cr }
\ee
and
\be
K=\pmatrix{\epsilon_{\alpha\beta}\otimes b_iV^i_{a_1a_2} &
m_{\alpha}^{\beta}\otimes \delta_{a_1}^{a_2}\cr
m_{\alpha}^{\beta}\otimes \delta_{a_1}^{a_2}&
\epsilon_{\alpha\beta}\otimes {\tilde b}_i{\tilde V}^i_{a_1a_2} \cr
}
~~~~~~~~
K^T=K~~~~\epsilon=\pmatrix{0&1\cr-1&0\cr}~~.
\ee
Then, the tree level superpotential \dash the matrix model potential \dash
can be written as:
\bea
W_{\it tree}&=&{1\over 2}{\cal Q}^TK{\cal Q}+{\cal Q}^T\cdot{\cal V}\nonumber\\
&=&{1\over 2}\left({\cal Q}+K^{-1}{\cal V}\right)^T K\left({\cal Q}+K^{-1}{\cal V}\right)
-{1\over 2}{\cal V}^TK^{-1}{\cal V}~~.
\eea
Therefore, the partition function is
\be
Z=\int DQ^{\hat \imath} D{\tilde Q}_{\hat \jmath}
\delta({Q_{\hat \imath}^\alpha{\tilde Q}_\alpha^{\hat \jmath}-
{\hat M}_{\hat \imath}^{\hat \jmath}}) e^{{1\over 2}{\cal V}^TK^{-1}{\cal V}
-{1\over 2}\ln\det K
}\Big|_{\it planar+1\,boundary}{}^{}~~.
\ee

The exponent of the integrand can be easily analyzed; fairly standard matrix
manipulations lead to:
\be
\det K=\left[\det{}_c\left(\delta_a^b\det m +
b_{\hat \imath}{\tilde b}^{\hat \jmath}
V^{\hat \imath}_{ac}{\tilde V}_{\hat \jmath}^{cb}
\right)\right]^2 ,
\ee
where $\det{}_c$ denotes a determinant over the color indices $a, b$,
while introducing the notation $V\equiv b_i V^i_{ab}$ and similarly for ${\tilde V}$,
the inverse of $K$ is given by:
\be
K^{-1}=\pmatrix{
-\epsilon\otimes {\tilde V}(\id_c \det m+V{\tilde V})^{-1}&
m^{-1}\otimes (\id_c +{{\tilde V}V\over \det m})^{-1}\cr
m^{-1}\otimes (\id_c+{V{\tilde V}\over \det m} )^{-1}&
-\epsilon\otimes {V}(\id_c \det m+{\tilde V}{V})^{-1}
}~~.
\ee

Let us now analyze in some detail the combination $(V{\tilde V}+\id_c \det m)^{-1}$.
Equation (\ref{zero})
implies that we need to compute only the terms proportional to the identity matrix.
The other terms will vanish upon contracting with ${\cal V}$.
It is not hard to see that
\be
(V{\tilde V})_a^b=-b{\hat M}^{-1}{\tilde b}\det{\hat M}\delta_a^b
+X_i^j({\hat M})Q_i^a{\tilde Q}^j_b,
\label{vtv}
\ee
where we have already used the $\delta$-function constraint from the path integral
to replace $Q_{\hat \imath}^a{\tilde Q}^{\hat \jmath}_a$ with
${\hat M}_{\hat \imath}^{\hat \jmath}$. This in turn implies that
\be
\left[(V{\tilde V}+\det m)^{-1}\right]{}_a^b={\delta_a^b
\over \det m-(b{\hat M}^{-1}{\tilde b})\det{\hat M}}+
Y^{\hat \imath}_{\hat \jmath}Q_{\hat \imath}^a{\tilde Q}^{\hat \jmath}_b~~.
\ee
The precise value of $Y$ is irrelevant, since the last term
always cancels due to contractions with $V_a$ or ${\tilde V}^b$

Thus
\bea
{\cal V}^TK^{-1}{\cal V}={2   (b  m{\tilde b})
\over \det  m-(b{\hat M}^{-1}{\tilde b})\det{\hat M}}
{\tilde V}^a V_a
=
{2(b m{\tilde b}) \det {\hat M}
\over \det m-(b{\hat M}^{-1}{\tilde b})\det{\hat M}} .
\eea

Combining all pieces together
we find that the gauge theory effective superpotential is given by:
\bea
\label{general}
-W_{\it eff}&=&-S\left(1-\ln {S\over \Lambda^3}\right)+S\ln{\det {\hat M}\over
\Lambda^{2(N_c-1)}}\\
&+&{(b m {\tilde b}) \det {\hat M}
\over \det m-(b{\hat M}^{-1}{\tilde b})\det{\hat M}}-
\ln\det\left(\det m\delta_a^b+
b_{\hat \imath}{\tilde b}^{\hat \jmath}
V^{\hat \imath}_{ac}{\tilde V}_{\hat \jmath}^{cb}\right).
\nonumber
\eea
The unit coefficient in front of the Veneziano-Yankielowicz term arises as the difference
between the number of gauge theory colors $N_c$ and the number of massless flavor
fields $N_f$.  Before we proceed, let us point out that the last term in the equation above
has an implicit dependence on the glueball superfield. Indeed, as the determinant is
taken over the matrix model color indices, the argument of the logarithm is of the
order $m^{2S}$. Exposing the part arising from the relevant planar diagrams
is potentially complicated; we will return to it shortly, after gaining some confidence
in the power of the matrix model.

\subsection{Comparison for $b_{\hat \imath}=0={\tilde b}^{\hat \imath}$}

Under this assumption the last
term in equation (\ref{general}) simplifies considerably, and the
effective superpotential reduces to:
\bea
W_{\it eff}=S\left(1-\ln {S\over \Lambda^3}\right)-
S\ln{\det {\hat M}\over \Lambda^{2(N_c-1)}}
-{(bm{\tilde b})\det {\hat M}
\over \det m}+S\ln {\det m\over \Lambda^2}.
\eea
To compare with the gauge theory effective superpotential we must integrate out
the glueball superfield.
\be
{\delta W_{\it eff}\over\delta S}=0~~~\Rightarrow~~~~{S\over \Lambda \det m }=
{\Lambda^{2(N_c-1)}\over \det {\hat M}}
\ee
and thus the effective superpotential is given by:
\be
W_{\it eff}={\Lambda^{2N_c-1}\det m \over \det {\hat M}}
-{(bm{\tilde b})\over \det m}\det {\hat M}~~,
\ee
which reproduces the field theory result. The first term can be recognized as the ADS
superpotential upon noticing that $\Lambda^{2N_c-1}\det m$ is the scale of the theory
obtained from the initial one by integrating out two quarks with mass matrix $m$.

\subsection{General analysis}

We now turn to analyzing the last term in equation (\ref{general}) and isolating
the part arising from planar and single boundary (in the sense of \cite{brr}) diagrams.

It is easy to reorganize this term using equation (\ref{vtv}).
To avoid cluttering the equations, let us introduce
\be
A=\det m-(b{\hat M}^{-1}{\tilde b})\det {\hat M}~~.
\ee
Then the last term in  (\ref{vtv}) becomes:
\bea
\ln\det\left(\delta_a^b\det m +
b_{\hat \imath}{\tilde b}^{\hat \jmath}
V^{\hat \imath}_{ac}{\tilde V}_{\hat \jmath}^{cb}\right)\!\!&=&\!\!
\Tr \ln A\delta_a^b+\Tr \ln \left[\delta_a^b+{1\over A}
X^{\hat \jmath}_{\hat \imath}Q_{\hat\jmath}^b{\tilde Q}^{\hat\imath}_a\right]
\nonumber\\
&=&\!\!S\ln A +\ln\det\!{}_f\left[\delta_{\hat \imath}^{\hat \jmath}
+{1\over A}
X^{\hat \jmath}_{\hat k}M^{\hat k}_{\hat\imath}\right]
\label{fudge}
\eea
where, as before, we identified the 't~Hooft coupling with the glueball superfield and
the matrix whose determinant is computed in the second term carries flavor indices.

We must now identify the leading terms in this equation - terms generated by
planar diagrams with as many gauge index loops as the diagrams with one boundary.
For this purpose it is important to notice that the computations in the previous
section yield the sum of {\it all} 1-loop n-point functions. Furthermore,
the planar, 1-boundary contribution must be proportional to the
number of colors $N_c\equiv S$, since there is one gauge index loop in such diagrams.
It is therefore clear that only the first term in equation (\ref{fudge}) should be kept since
the determinant in the second term is in flavor space and there is no term in
its expansion which is proportional to the number of colors. Thus, the gauge theory
effective superpotential is given by:
\bea
W_{\it eff}&=&S\left(1-\ln {S\over \Lambda^3}\right)-S\ln{\det {\hat M}\over
\Lambda^{2(N_c-1)}}\label{Weff(s)}\\
&-&
{(bm{\tilde b})\det {\hat M}
\over \det m -(b{\hat M}^{-1}{\tilde b})\det{\hat M}}+
S\ln {1\over \Lambda^2}\left(\det m -(b{\hat M}^{-1}{\tilde b})\det {\hat M}\right).
\nonumber
\eea

Integrating out $S$ leads to:
\be
{S\over \Lambda^3}={\Lambda^{2(N_c-2)}}\left[
{ \det m \over \det {\hat M}}-(b{\hat M}^{-1}{\tilde b})\right]
\ee
which in turns implies that the effective superpotential is:
\bea
W_{\it eff}&=&{\Lambda^{2N_c-1}}\left[
{ \det m\over \det {\hat M}}-(b{\hat M}^{-1}{\tilde b})\right]
-
{(b m {\tilde b})\det {\hat M}
\over \det m-(b{\hat M}^{-1}{\tilde b})\det{\hat M}}
\label{finrez_mm}
\eea
This reproduces the field theory result (\ref{ft}, \ref{ft2}).


\section{Vacua \label{general_all}}

Although it is already clear that there is an exact agreement between the matrix model
and gauge theory, let us briefly discuss the vacua of the gauge theory and their
construction from the matrix model. In gauge theory we need to integrate out all
mesons and
baryons, while on the matrix model side we need to compute the full partition
function. We begin with the gauge theory discussion. We will discuss the
construction in the language of section \ref{all_out} and relate it at the end with
section \ref{hard}.


\subsection{Integrating out all fields in gauge theory\label{generalft_all}}

Let us recall equation ~(\ref{geneq_all}), which determines the low energy effective 
superpotential for a general $N_f\ne 3$:
\begin{displaymath}
  (N_f-1)^2 x^{-1/(N_f-1)} x f'(x)
  [ f(x) + (N_f-1) (N_f-3) x f'(x) ]
  = - 1.
\end{displaymath}
Besides this equation, there are other equations $f(x)$ obeys,
obtained from varying the dynamical superpotential with respect to the mesons:
\begin{equation}
  \frac{B^i \tilde{B}_{{j}} - (\det  M)
    (M^{-1})^{i}_{{j}}}{\Lambda^{2N_c-1}} + m^i_{j} = 0.
\end{equation}
Using various equations from section 3.1 this equation can written as:
\begin{eqnarray}
  \lefteqn{
    B^i \tilde{B}_{{j}} - (\det  M) (M^{-1})^{i}_{{j}}
     = -m^i_{j} \Lambda^{2N_c-1}} \nonumber \\
  &=& \frac{(m\tilde{b})^i (bm)_{{j}}}{(bm\tilde{b})}
  (N_f-1)^2 \biggl[ (N_f-1)^2 xf'(x)
  [(\det m) \Lambda^{2N_c-1}]^{2/(N_f-1)} \frac{xf'(x)}{(bm\tilde{b})}
  \nonumber \\
  & &\!\!\!\!\!\!\!\!\!\!\!\!\!\!\!\!\!
    + \Lambda^{2N_c-1} (f(x)-2(N_f-1)xf'(x))^{N_f-2}
    (f(x)+(N_f-1)(N_f-3)xf'(x)) \biggr]
  \nonumber \\
  & & \!\!\!\!\!\!\!\!\!\!\!\!\!\!\!\!\!
- m^i_{j} \Lambda^{2N_c-1}
  (f(x)-2(N_f-1)xf'(x))^{N_f-2} \left[f(x)+(N_f-1)(N_f-3)xf'(x)\right]. \nonumber \\
\end{eqnarray}
To satisfy this equation, the coefficient of $(m\tilde{b})_i
(bm)_{{j}}$ in the square bracket must vanish, and the
coefficient of $m^i_{j}$ must agree on both sides.  Therefore
we find
\begin{eqnarray}
\!\!\!\!
(N_f-1)^2 x^{-1/(N_f-1)} xf'(x) + (f(x)-2(N_f-1)xf'(x))^{N_f-2} \!\!&=&\!\!
  0~,~~\\
\!\!\!\!
(f(x)-2(N_f-1)xf'(x))^{N_f-2} (f(x)+(N_f-1)(N_f-3)xf'(x)) \!\!&=&\!\! 1~.~ \label{last}
\end{eqnarray}

Thus, there seem to be three equations for a single function; it turns out
however that one of them can be obtained from the other two.
In general, we cannot expect to find a consistent solution for two
first-order differential equations for one function. In the $N_f=3$ case, the two differential
equations were self-consistent, and their combined effect was to fix the integration constant
in $f(x)$ \footnote{Indeed, if one did not fix the integration constant $c$ in
$f(x) = \pm \sqrt{c-x^{1/2}}$ using
the boundary conditions, equation (\ref{last}):
\begin{eqnarray}
  \left(\sqrt{c-x^{1/2}} -4 x \frac{-\frac{1}{2}x^{-1/2}}{2\sqrt{c-x^{1/2}}}
    \right) \sqrt{c-x^{1/2}} = 1,
\end{eqnarray}
would fix this constant to be $c=1$.}. We expect the same to happen here.

In general, we can solve for $f'(x)$ using Eq.~(\ref{geneq_all}), and
substitute it to one of the other equations.  Since
Eq.~(\ref{geneq_all}) is quadratic in $f'(x)$, it has two
solutions. Only one of them is consistent with the
boundary condition $|f(0)|=1$.  Keeping only the consistent solution, we find
\begin{eqnarray}
  & &\left( \frac{(N_f-1)(N_f-2) f
      - \sqrt{(N_f-1)^2 f^2 - 4(N_f-1)(N_f-3) z }}
    {(N_f-3)(N_f-1)} \right)^{N_f-2}
  \nonumber \\
  & &
  + \frac{z^{-1}
    \left( - (N_f-1) f +
      \sqrt{(N_f-1)^2 f^2 - 4(N_f-1)(N_f-3) z } \right) }{2 ( N_f -3 )}
  = 0,
\label{hit}
\end{eqnarray}
where $z = x^{1/(N_f-1)}$.
This equation determines the function $f(x)$ implicitly.


The same results can be obtained following the steps in section \ref{hard}. In particular,
when all $N_f=N_c+1$ flavors are integrated out, equations (\ref{det}) and (\ref{e1}) become:
\bea
\det M &=& - B X -{1 \over B} \\
\det M^{N_f-1} + X B \det M^{N_f-2} &=& \det m ~ \Lambda^{N_f(2N_c-1)}
\eea
The effective superpotential is then obtained by substituting the solutions of these
equations in the superpotential (\ref{weff})
\be
\Lambda^{2Nc-1} W_{eff}= (B^2 X+ N_f -1 ) \det M + B^3 X^2 + 3 B X.
\ee
It is not hard to check that this reproduces the results in chapter 3
for the case of an $SU(2)$ gauge group with 3 flavors; it is, however, somewhat more
challenging to see that it agrees with (\ref{hit}) as well.

\subsection{The Matrix Model Free Energy\label{mm_all}}

Let us now consider the matrix integral we considered before, but with all flavors
massive. In this case, we can reinterpret the $\delta$-function as arising from the
change of variables
\be
\int DQD{\tilde Q} =  \int D M \int DQD{\tilde Q}\delta(Q{\tilde Q}-M)
\ee
Thus, to find the effective superpotential as a function of the glueball superfield
(\ref{DV}), we must supplement the results of the previous section with a mass term
for the remaining mesons and then compute the integral over $M$ as well.
We recall that we are interested only in the 1-boundary free energy. Thus, the
integral can be computed by a saddle-point approximation. Alternatively, it is easy to see
that the one-boundary free energy is given by the sum of all tree-level Feynman
diagrams arising from the superpotential (\ref{Weff(s)}). This implies that,
as expected, the effective superpotential is unique even when expressed in terms of
the glueball superfield. The vacua of the theory arise in this language as the critical
points of $W_{\it eff}(S)$.

We now illustrate this simple observation for the $SU(2)$ theory with three flavors,
leaving to the reader the exercise of recovering the more involved results of
section \ref{generalft_all}.

\subsection{Back to $SU(2)$}

Consider the equation (\ref{Weff(s)}) for the case of an $SU(2)$ theory with three
flavors. Since ${\hat M}$ is 1-dimensional, the superpotential is:
\bea
W_{\it eff}&=&S\left(1-\ln {S\over \Lambda^3}\right)-S\ln{ {\hat M}\over
\Lambda^{2}}  \\
&-&
{(bm{\tilde b}) {\hat M}
\over \det m -(b_l{\tilde b}_l)}+
S\ln {1\over \Lambda^2}\left(\det m -(b_l{\tilde b}_l)\right)+m{\hat M}
\nonumber
\eea
where $b_l$ and ${\tilde b}_l$ are the sources with indices along the meson which
was not integrated out in the previous section. The saddle point equation is:
\be
{S\over {\hat M}}=m-{(bm{\tilde b})
\over \det m -(b_l{\tilde b}_l)}={\det {\check m}-(b{\check m}{\tilde b})
\over \det m -(b_l{\tilde b}_l)}
~~~~{\rm where}~~~~{\check m} =\pmatrix{ m_\alpha^\beta&0\cr 0 &m\cr}
\ee
and ${b}$ and ${\tilde b}$ are understood as 3-component vectors.
Then, the effective superpotential as a function of the glueball superfield is:
\bea
W_{\it eff}(S)&=&2S\left(1-\ln {S\over \Lambda^3}\right)-S\ln{\Lambda^3\over
\det {\check m}-(b{\check m}{\tilde b})}
\eea

As argued before, the vacua are now described by  the critical points of $W_{\it eff}(S)$, and
are given by
\be
2\ln {S\over \Lambda^3}=\ln{\det {\check m}-(b{\check m}{\tilde b})\over \Lambda^3}
~~~~\Leftrightarrow~~~~
S=\pm \sqrt{\det {\check m}-(b{\check m}{\tilde b})}\,\Lambda^{3/2}
\ee
The superpotential at the critical points is therefore:
\be
W_{\it eff}\Big|_{\it crit}=\pm 2\sqrt{\det {\check m}-(b{\check m}{\tilde b})}
\Lambda^{3/2}~.
\ee
We thus recover the matrix model result found directly in equation (\ref{finegsu2_1})
of section \ref{mm2},  as well as the field theory result.


\section{Baryons and Geometric Transitions}

In this section we discuss the baryons in the context of the geometric
transitions. The gauge theory is engineered by wrapping  D5 branes
on several compact $\P^1$ cycles of a geometry which locally, around
each cycle, is the geometry of the small resolution of the conifold. Alternatively,
it can be described using the T-dual brane configuration, where the D5 branes
wrapped on $\P^1$ cycles are mapped into D4 branes stretched between
NS branes \cite{rtw}, \cite{dot1,dot2,dot3}.

Let us begin by briefly reviewing the results of \cite{OOHO,GIKU},
concerning the baryonic degrees of freedom in MQCD. First, we need to
comment on having an $SU(N)$ rather than an $U(N)$ gauge group. The Type IIA
brane configuration as well as the Type IIB geometric construction
describe a classical $U(N)$ gauge theory. The M theory limit describes a
quantum $SU(N)$, where the $U(1)$ factor decouples.  As explained in
\cite{dot1}, the $U(1)$ factor is recovered after the geometric
transition, when the $SU(N)$ part confines.
Therefore, the approach of \cite{rtw} cannot be applied for the case of
baryons, as the quantities in matrix models were obtained from the parameters of
brane configurations via lifting to M theory.

It is nevertheless possible to collect
some information about the vacuum expectation values
of the baryon operators in MQCD.
As described in \cite{OOHO}, in the case $N_f = N_c$, the difference between
a baryonic and a non-baryonic branch is that the
asymptotic regions of the former intersect, and the ones of the latter do not.
Indeed, the
asymptotic regions for the non-baryonic branch are given by:
\begin{eqnarray}
t &=& (w^2 + \Lambda_{N=1}^4)^{N_c/2}~~,~~~v = 0 \nonumber \\
t &=& \Lambda_{N=1}^{2 N_c}~~~~~~~~~~~~~~~~,~~~w = 0
\label{nbar}
\end{eqnarray}
while the ones for the baryonic branch are given by:
\begin{eqnarray}
t &=& w^{2N_c}~~~,~~~v = 0  \nonumber\\
t &=& \Lambda_{N=1}^{2 N_c}~~,~~~w = 0 ~~.
\label{bar}
\end{eqnarray}
It is clear that the two branches intersect in  (\ref{nbar}),  but are
separated in (\ref{bar}).
The distance between the asymptotic regions in (\ref{bar}) is the value of
$\tilde{B} B$.

In M theory terms  the geometric transition corresponds to
a transition from an M5 brane with a worldvolume
containing a Riemann surface in the $(v,w,t)$ plane to an M5 brane with two
dimensions
embedded in $(v, w)$, for constant $t$. The equation in $(v, w)$ represents an NS brane
which is T-dual to the deformed conifold.

In the case of (\ref{nbar})-(\ref{bar}), $v$ and $w$ are decoupled
so the above discussion does not apply.
In the language of \cite{rtw}, this can be understood by starting with D4$_m$
branes corresponding to massive flavors, taking the mass to zero and combining with a
color D4 brane to get a D4$_M$ brane
which describes a flavor with an expectation value.  Therefore, in the geometrical
picture, there are no D5 branes on the compact $\P^1$ cycles and there are
only D5 branes on the noncompact 2-cycles.
We then see that the duality between matrix models and field theory fails in this case.

The only way to use the results of \cite{dv1,dv2,dv3} is to give mass to one
of the flavors, which means decomposing one D4$_M$ brane into a D4$_m$ brane and
a color brane. This is exactly the procedure
discussed in detail in \cite{brr} where a method to deal with this case was
stated.
Therefore, we see that the difficulties with the matrix model analysis of the
baryon operators have a geometric counterpart. This should probably be expected,
since the geometry is underlying the matrix models.

\section{Conclusions}

In this paper we further analyzed the extension of the Dijkgraaf-Vafa proposal
to theories containing fields in the fundamental representation. While this extension
was thoroughly analyzed in situations in which the gauge theory was
described solely in terms of mesons, the matrix model description of
baryonic deformation remained until now largely unexplored.
The main goal of our work was to fill this gap.

We have started with the ${\cal N} = 1$ SQCD with gauge group $SU(N_c)$ and
$N_c + 1$ flavors  whose effective superpotential was conjectured in \cite{seiberg} and
deformed the theory by adding baryon sources as well as mass terms for either two or
all flavor fields. We compared the resulting effective superpotential obtained by
integrating out the appropriate mesons and baryons with the one coming
from the matrix model computations and we found perfect agreement.
Of essential importance has been the correct identification of Feynman diagrams
contributing to the superpotential.


We expect that the effective superpotential for other s-confining theories is
computable using matrix model techniques along the lines described here,
after suitable deformations by mass terms and other sources.

SQCD theories with $N_f \ge N_c+2$ are usually analyzed using Seiberg's duality.
One may ask whether the matrix model techniques can shed light on their
effective superpotential. Using 't~Hooft's anomaly matching conditions it was
shown that the mesons and baryons are not the only low energy degrees of freedom.
However, the complete set of low energy fields is not known. Nevertheless, by inserting
sources for the known fields in the tree level superpotential, the matrix model
perturbation theory should allow one to recover the truncation of the full effective
superpotential to these fields.

\medskip

{\bf Acknowledgements}

\vskip 3mm
We thank Per Kraus and Eric d'Hoker for useful discussions. R.R.
thanks the Michigan Center for Theoretical Physics for hospitality. 
The work of I.B. was supported by the NSF under
Grants No. PHY00-99590, PHY98-19686 and PHY01-40151.
H.M. and R. T. are supported in part by the DOE Contract
DE-AC03-76SF00098 and in part by the NSF grant PHY-0098840.
The work of R.R. was supported in part by DOE under Grant No. 91ER40618
and in part by the NSF under Grant No. PHY00-98395.


\end{document}